# Inelastic Effects in Low-Energy Electron Reflectivity of Two-dimensional Materials

Qin Gao, P. C. Mende, M. Widom, R. M. Feenstra
Dept. Physics, Carnegie Mellon University, Pittsburgh, PA 15213

## Abstract

A simple method is proposed for inclusion of inelastic effects (electron absorption) in computations of low-energy electron reflectivity (LEER) spectra. The theoretical spectra are formulated by matching of electron wavefunctions obtained from first-principles computations in a repeated vacuum-slab-vacuum geometry. Inelastic effects are included by allowing these states to decay in time in accordance with an imaginary term in the potential of the slab, and by mixing of the slab states in accordance with the same type of distribution as occurs in a free-electron model. LEER spectra are computed for various two-dimensional materials, including free-standing multilayer graphene, graphene on copper substrates, and hexagonal boron nitride (h-BN) on cobalt substrates.

## I. Introduction

For decades, low-energy electrons (0 – 300 eV) have been employed as a probe of the geometric and electronic structure of surfaces. Since these electrons interact very strongly with atoms, any electrons that are elastically scattered from the surface necessarily originate from only the top few surface layers, hence providing a sensitive probe of the near-surface region. In particular, the technique of low-energy electron diffraction (LEED) has been developed by many workers, both experimentally and theoretically.[1,2] In a conventional LEED instrument, a mono-energetic electron beam is directed towards a surface, and the pattern of diffracted beams is recorded. Each diffracted beam can be labeled by one (or more) reciprocal lattice vector(s), $\mathbf{g} = (g_x, g_y)$. These diffraction vectors are often denoted by two integer or fractional values ($p,q$) such that $\mathbf{g} = p\mathbf{b}_1 + q\mathbf{b}_2$ where $\mathbf{b}_1$ and $\mathbf{b}_2$ are the basis vectors of the reciprocal lattice of the projected bulk structure.[3] The intensity of the diffracted beams as a function of incident energy, $I_\mathbf{g}(E)$, can be measured. Such $I_\mathbf{g}(E)$ curves contain valuable information on the geometric structure of the surface, in particular for cases when the surface is *reconstructed* giving rise to fractional $p$ and $q$ values. However, in most cases the actual atomic positions are very difficult to directly deduce from simple inspection of the $I_\mathbf{g}(E)$ curves, so that it is necessary to perform a detailed comparison between experimentally measured and theoretically predicted curves in order to determine the structure.[1,2,3]

The development of a theory by which LEED $I_\mathbf{g}(E)$ curves could be computed is a problem that was intensively studied during the 1970s and 1980s.[1,2] A method for addressing the problem was eventually developed in which multiple scattering of the incident electrons was included. This procedure entailed first considering scattering by individual atoms, then by a collection of atoms forming a layer, and finally by a collection of layers to form the solid. By about 1990 the theoretical procedures (and computer implementation) existed whereby the $I_{\mathbf{g}}(E)$ curves could be computed with a fair degree of confidence, although the theory was not reliable for energies < 50 eV. One reason for this inaccuracy at energies < 50 eV is that the detailed electronic structure of the solid (e.g. formation of bands) is, in fact, largely untreated in the theory. Since that time, two important developments in low-energy electron research have occurred: First, the development of the low-energy electron microscope (LEEM) has enabled measurement of $I_{\mathbf{g}}(E)$ curves over energies of typically 0 – 50 eV, particularly for the (0,0) nondiffracted beam, $I_0(E)$.[4] Such measurements were performed even prior to the LEEM,[5] but they greatly gained in popularity with the advent of the LEEM.[6,7,8,9,10,11,12] Second, computational methods for obtaining the electronic band structure of solids have greatly advanced (i.e. not specifically for LEED), with the eventual establishment for public-domain computer codes such as the Vienna Ab Initio Simulation Package (VASP)[13,14,15] which have large numbers of experienced users and for which the packages experience continual improvements (e.g. with new pseudopotentials and new density functionals).

With these developments, we undertook a project two years ago in which a theoretical description of LEER spectra at very low energies of 0 – 20 eV was formulated, employing first-principles wavefunctions obtained from VASP.[12,16,17] Linear combinations of the VASP pseudo-wavefunctions were formed, such that they matched incoming or outgoing waves in the vacuum (or the substrate). At the very low energies, good results for LEER spectra were obtained for the case of multilayer graphene films, both free-standing or on a metal substrate such as copper. However, despite this success, several limitations in the theory remained: (i) it did not incorporate inelastic effects (electron absorption); (ii) for a film on a substrate, only simple metals having free-electron-like states at the relevant energies could be dealt with; and (iii) the theory for diffracted beams was not developed.

In this work, we present a model for addressing the first of these limitations regarding electron absorption. In the model, an *imaginary* component to the potential in the slab is introduced, as in past LEED theory. Then, considering wavepackets centered about each of the VASP wavefunctions, a phase-shift analysis on the reflected wave is employed in order to evaluate the time spent in the slab by the wavepacket. From that time, and using the imaginary component of the potential, an attenuation factor for each of the VASP states is obtained. Then, considering the spatial exponential decay of the actual electron wavefunctions in the near-surface region, a distribution of the VASP states is formed to permit that decay (with the form of the distribution taken from a free-electron model for the states in the slab). With this distribution, the final values for the electron reflectivity *including* absorption are evaluated by summing over the states, weighting each term by its known attenuation factor.

It is important to note that a rigorous theoretical method for incorporating inelastic effects in very low-energy LEER spectra already exists, from the work of Krasovskii and co-workers.[18,19,20] Their methodology goes well beyond the earlier LEED analysis procedures in that it addresses the detailed band structure of the solid. It also includes electron attenuation (dealt with, again, by using an imaginary component of the potential) as well as electron states that decay spatially into the material, i.e. evanescent states with imaginary wavevector values. The theory of Krasovskii et al. differs from our own simulation procedure for LEER spectra in that it more exactly incorporates these evanescent (imaginary wavevector) and inelastic (imaginary potential) terms in a complete "inverse band-structure" formulation of the problem. Their results provide a benchmark against which we can compare our approximate results.

We emphasize that we have not in any way incorporated the imaginary term of the potential directly into the VASP code. Such an approach would lead to non-Hermitian matrices within VASP, which would require very significant modification of the code to handle. Rather, our approach is purely based on post-processing of the VASP eigenstates. We treat these states first by neglecting any inelastic effects and forming the electron reflectivity as described in our prior works.[12,16,17] We then incorporate the imaginary part of the potential in a two-step procedure, considering both the time-decay of individual eigenstates as well as the mixing of eigenstates with ones nearby in energy.

This paper is organized as follows. In the next Section we provide a detailed description of the model that we use for incorporating inelastic effects. In Section III, we provide computational results for multilayer graphene, both free-standing and on copper. Absorption effects are found to be small in those cases. We also consider the case of h-BN on cobalt substrates, for which electron absorption effects are found to be large (due to the band structure of the h-BN films). In general, our computed results compare well with published experimental LEER spectra for the various materials, and with our computational results we are able to provide improved understanding of those spectra. Finally, our work is summarized in Section IV.

## II. Theoretical Methodology
### A. Reflectivity Computations without Inelastic Effects

Our method for obtaining the reflectivity, $I_0(E)$ normalized to the incident beam current, has been previously described.[12,16,17] Briefly, we first consider a periodically repeated vacuum-slab-vacuum system, with the slab consisting typically of several graphene layers (for computing reflectivity of free-standing multilayer graphene), or several layers of a substrate material covered on one or both sides by layer(s) of some overlayer material (for computing reflectivity of the overlayer on the substrate). In both cases, VASP computations are performed to obtain the eigenvalues and eigenstates of the system, usually performing computations for a few different widths of the vacuum layer. For the case of free-standing graphene, linear combinations of these VASP wavefunctions are formed such that there is only an outgoing wave on the right-hand side of the slab. On the left-hand side there then exist both incoming and reflected waves, and by taking the ratio of their respective amplitudes the reflectivity is obtained. For the case of an overlayer on a substrate, the methodology also requires a VASP computation of the bulk band structure of the periodically repeated substrate material. For the simple-metal substrates considered to date, they are found to have nearly-free-electron (NFE) states located above the vacuum energy, and linear combinations of the slab states are formed to match these NFE states propagating into the substrate. Then, by examination of the incident and reflected waves on the left-hand side of the slab the reflectivity is deduced. Our method used in the present work is identical to that of our previous work, except for one technical detail having to do with the manner in which we reject states of "mixed" character (which act to couple slabs in adjoining simulation cells) from the analysis; Appendix I describes an improved method for dealing with such states.

## B. Model for Inelastic Effects

We now consider the inclusion of inelastic effects in the reflectivity computation, something that was not considered in our prior work. Consider a wavepacket centered about some energy *E*, incident on a surface. In LEER measurements we are concerned with the wavepacket that is reflected from the surface. To include inelastic effects in the reflectivity, we employ a model consisting of two parts. In the first part, we perform a rigorous computation of the scattering phase shift of the reflected wave, and from that we can deduce both the travel time for the reflected wavepacket and its dwell time in the slab. This underlying basis for this analysis is well described by Merzbacher.[21] Using our previous notation,[12] we consider a simulation cell extending over $-z_S \leq z \leq z_S$, with our slab of material in the center of this cell. We evaluate the ratio of the reflected to incident wavefunction amplitudes at the far left-hand of the cell, $z = -z_S$. The magnitude squared of that complex quantity gives us the reflectivity, as described previously.[12,17] We then employ the phase $\phi$ of this ratio, corresponding to the phase shift between incident and reflected waves. As discussed by Merzbacher, the round trip time of a wavepacket from $z = -z_S$, to (and within) the slab, and back to $z = -z_S$, is given simply by $\hbar(d\phi/dE)$. In principle, the phase obtained at each energy is known only modulo $2\pi$ (and an additional uncertainty of $\pi$ arises since the reflected wave has a $\pm\pi/2$ phase shift as detailed in Ref. [21]). However, by carefully examining the results as a function of energy, the requisite factors of $\pi$ can be inserted such that a smooth $\phi(E)$ curve is obtained.

To obtain a dwell time for the wavepacket within the slab, we must subtract the time spent traveling in the vacuum. For this purpose we must define an edge of the slab, which we denote by $z = -z_E$. For example, we can take the left-most atomic layer of the slab and define $-z_E$ to be at one atomic radius farther to the left of that plane. In order to minimize any influence of this choice of $-z_E$ on our resulting dwell time, we subtract the travel time in the vacuum *including* the presence of the potential (averaged over the *x* and *y* directions) in the vacuum. We denote that potential by $V(z)$ with $V(z) \to 0$ for $z \ll 0$ and $V(z) < 0$ everywhere. For a state with energy *E,* a semi-classical velocity is given by $\sqrt{2[E-V(z)]/m}$, and the travel time is obtained by integrating the inverse of this quantity over the range $z = -z_S$ to $-z_E$. Thus, our final expression for the dwell time in the slab for a state $\alpha$ with energy $E_\alpha$ is given by

$$\Delta t_\alpha = \hbar \left(\frac{d\phi}{dE}\right)_{E_\alpha} - 2\int_{-z_S}^{-z_E} \frac{dz}{\sqrt{2[E_\alpha - V(Z)]/m}} \quad (1)$$

with the derivative evaluated at $E = E_\alpha$ and where the factor of 2 before the integral sign arises since we require the round trip travel time in the vacuum. With this definition, we find that resultant values for $\Delta t_\alpha$ are only weakly dependent on our choice of $-z_E$, as demonstrated in the following Section.

As discussed by Merzbacher, the dwell time is large when the energy of the incident wave corresponds to the energy of a state in the slab. Such states are all resonances, of course, since they are degenerate with the continuum of propagating states in the vacuum. When a particular resonant state has a narrow energy spread, it can produce a long dwell time, i.e. the incident electron spends considerable time "bouncing to and fro" in the slab before being reflected.[21] Given the dwell time, we then assume some imaginary component of the potential in the slab, $-iV_i$ with $V_i > 0$, and we thus arrive at an attenuation factor $\exp(-\Delta t_\alpha / \tau)$ where a characteristic decay time is given by $\tau = \hbar / 2V_i$ (the factor of 2 in the denominator arising since we are considering the magnitude squared of the wavefunction).[22]

Having obtained this attentuation factor as a function of energy, we could in principle use that to compute a reflectivity spectrum. We would simply correct the elastically-computed reflectivity at each energy by its respective attentuation factors. However, comparing such results to experiment, we find poor agreement in certain cases. In particular, when a sharp resonance with its concomitant long dwell time produces strong attenuation, it is found experimentally that this attentuation *extends out to neighboring energies*, over a range of > 1 eV. Such effects are clearly apparent in the rigorous computations of Krasovskii et al., e.g. Fig. 2 of Ref. [18], in which the strong attenuation associated with narrow elastically-computed reflectivity minima is seen to broaden out to neighboring energies.

The reason for this broadening or smearing of the attentuation is clear if we reconsider the inelastic (electron-electron) interactions. The incident wavepacket, centered at an energy $E$, will be *spatially* attenuated as it propagates into the solid.[22] This exponentially decaying wave in the solid will in general be described by a linear combination of the single-particle states. Hence, we can view the incident states as *mixing* with other states; this mixing is the second part of our methodology, and for this we employ a free-electron model to describe the distribution of mixed states. That is, we assume that this distribution is the same in the real system as it is for a free-electron model with the slab characterized simply by a constant inner potential, $-V_0$ where $V_0 > 0$. By combining these two parts of the model, first mixing the incident wavepackets into a distribution of states in the slab, and then attenuating each component of that mixture in accordance with its dwell time in the slab, we arrive at a complete (albeit phenomenological) model for the inelastic effects.

The relevant exponential decay length in the solid can be obtained by multiplying some velocity times the decay time $\hbar / V_i = 2\tau$. This velocity should, in principle, be the group velocity of our wavepacket. However, for the narrow resonances discussed above that group velocity is relatively small, corresponding to rather short (perhaps unphysical) decay lengths. Moreover, to decompose that spatially exponentially decaying state into single-particle (nondecaying) states of the slab would be a complex procedure. Therefore, at this point we employ a model based on a free-electron description of the states in the slab. In that case, the velocity is given simply by $\sqrt{2[E-V_0]/m}$ corresponding to a decay length of $\lambda_\alpha = 2\tau\sqrt{2[E_\alpha - V_0]/m}$ for energy $E_\alpha$. We consider a state in the slab with exponentially decaying form given by $\exp(ik_\alpha z)\exp(-z/\lambda_\alpha)$ for $z > 0$, where $k_\alpha = \sqrt{2m(E_\alpha + V_0)/h^2}$ is the wavevector in the $z$-direction. This form is Fourier analyzed to obtain the distribution of single-particle states labeled by $\beta$ that compose it, yielding a distribution with terms proportional to $1/[(k_\alpha - k_\beta)^2 + (1/\lambda_\alpha)^2]$ where $k_\beta = \sqrt{2m(E_\beta + V_0)/\hbar^2}$ . We take the square of these terms, i.e. as applicable to the wavefunction squared, yielding the distribution

$$F_{\alpha\beta} \equiv \frac{c_\alpha}{[(k_\alpha - k_\beta)^2 + (1/\lambda_\alpha)^2]^2} \qquad (2)$$

where the normalization factor $c_\alpha$ is set by the condition that $\sum_\beta F_{\alpha\beta} = 1$. With this distribution, inelastic effects are then computed by the following procedure: (i) for a given spectrum computed without inelastic effects, we have the reflectivities $R(E)$ and phases $\phi(E)$ from which we obtain the dwell times, $\Delta t_\alpha$, and the attenuation factors, $\exp(-\Delta t_\alpha/\tau)$, for a state $\alpha$; (ii) for each state we compute the distribution $F_{\alpha\beta}$; and (iii) the resultant reflectivity for the state is given by $\sum_\beta F_{\alpha\beta} R(E_\alpha) \exp(-\Delta t_\beta/\tau)$.

The above free-electron picture involves modeling the states of the semi-infinite solid by assuming a constant potential $-V_0$ in the solid, where $V_0 > 0$ is known as the *inner potential*. Inner potentials are commonly employed in approximate treatments of electron interactions with solids; they are deduced by a variety of means, yielding different results depending on the method and the energy range of the states of interest.[23,24] In our case we are interested in very low-energy electron states, typically 0 – 15 eV above the vacuum level or, equivalently, 5 – 20 eV above the Fermi level. To deduce an inner potential applicable to that energy range we examine the band structure of the material, matching that to a free-electron model with bands of the form

$$E_n = -V_0 + \frac{\hbar^2}{2m}|\mathbf{k} + \mathbf{G}_n|^2 \qquad (3)$$

where $\mathbf{k}$ is the electron wavevector with components $k_x, k_y, k_z$ all ranging from $-\infty$ to $+\infty$, and $\mathbf{G}_n$ are reciprocal lattice vectors (*n* being a band index) chosen such that $\mathbf{k} + \mathbf{G}_n$ falls within the first Brillouin zone (BZ).[25] We focus on the wavevector direction that is perpendicular to the surface we are considering; as part of our standard reflectivity analysis (i.e. even without inelastic effects) we have already computed, by first principles, band structures in that direction.

For example, consider the band structure for Cu in the (111) direction shown in Fig. 6 of Ref. [17]. A dispersive band intersects the edge of the BZ for an energy of 16.7 eV. We can match the bands near that energy with free-electron bands from Eq. (3), employing an inner potential of 7.4 eV relative to the *Fermi energy* and with reciprocal lattice vectors of 2 and $-3$ times that of the first reciprocal lattice vector for periodically repeated ABC-stacked Cu. (Other bands of the first-principles results are also approximately matched by the free-electron model, though less precisely.) Then, to deduce an inner potential relative to the *vacuum energy*, we match the computed potential for the bulk Cu with one consisting of a free-standing three-layer copper slab. From that we deduce a work-function for the Cu(111) surface of 4.7 eV, and thus the inner potential is 12.1 eV relative to the vacuum level. Of course, this work function might well change for different overlayers on Cu (or for different surface orientations), but as demonstrated in the next Section the results of our model for inelastic effects are very insensitive to the actual value of inner potential that we use.

As another example, consider the bulk bands of graphite in the (0001) direction, shown by Hibino et al. in Fig. 3 of Ref. [8] (we obtain nearly identical bands in our own computations using VASP, as pictured in Fig. 1(b)). There is a dispersive band intersecting the edge of the BZ at an energy of 7.0 eV above the Fermi energy. Again, modeling this band by free-electron states, we use Eq. (3) with an inner potential of 13.9 eV and reciprocal lattice vectors of 2 and −3 times that of the first reciprocal lattice vector for periodically repeated AB-stacked graphite. Then, comparing computed potentials of bulk graphite with a two-layer graphite slab, we deduce a work-function of 4.2 eV and hence an inner potential of 18.1 eV relative to the vacuum level. These results, 12.1 eV for Cu and 18.1 eV for graphite, are comparable to previously deduced inner potentials of 13.4 and 17 eV, respectively, for the two materials.[23,24] For situations in which we have one or a few layers of some material on a substrate, we then employ in our model simply the *average* inner potential of the overlayer and the substrate.

## III. Results
### A. Free-standing Graphene

In Fig. 1(a) we show the computed reflectivity for a free-standing 4-layer slab of multi-layer graphene, showing results both with and without inelastic effects. Oscillations are seen in the spectra in the energy ranges 0 – 6 and 14 – 22 eV. As argued by Hibino et al.,[8] these oscillations are associated with electronic states that derive from dispersive bands of graphite occurring over the same energy ranges, as shown by the orange-colored bands in Fig. 1(b) (the black bands there are also graphite-derived bands, but they do not have the appropriate character to couple to incident plane-wave states and so do not play any role in determining the reflectivity). Fundamentally, the reflectivity minima can be understood in terms of *interlayer states* between the graphene planes,[26] as explained in our prior work.[12] The results of Fig. 1(a) are compiled from computations using three different vacuum widths; some small gaps remain between the groups of reflectivity points, but all important features of the spectrum are clearly evident. The inner potential for the inelastic computations of Fig. 1 is taken to be 18.1 eV, and the edge of the slab is placed at one atomic radius of carbon, 0.7 Å, to the left of the left-most graphene plane.

For the imaginary part of the potential we assume a linear energy dependence, $V_i = 0.4\,\text{eV} + 0.06\,E$. This dependence is comparable to what is used in the prior work of Krasovskii et al.[18,19,20] As demonstrated by those authors, $V_i$ will in general increase monotonically, although sometimes in a stepwise manner as new channels for inelastic electron interactions appear as the energy increases. As a first approximation we employ this linear dependence of $V_i$ with energy; the same dependence was used by Flege et al. in Ref. [20]. For multi-layer graphene we have chosen the parameter values in this linear dependence such that the amplitudes of the reflectivity oscillations in the energy ranges 0 – 6 eV and 14 – 22 eV approximately match experiment.[9] We use the same parameters in our treatments of graphene on copper and h-BN on cobalt as described in the following Sections.

The lower energy range is displayed in more detail in Fig. 2, showing a comparison of theory with experimental results from various numbers of graphene layers on SiC (these results qualitatively resemble the situation for free-standing graphene, as explained in Ref. [12]).[8,9,11] The inelastic effects are seen to significantly diminish the peak-to-valley amplitude of the observed oscillations in the 0 – 6 eV range, to a peak-to-valley reflectivity variation of ≈0.05 for the 6 monolayer (ML) case. We note that the theoretical prediction in Fig. 2 for this peak-to-valley variation is somewhat too large for the 2 – 4 ML cases. A likely reason for this discrepancy is that, theoretically, we are modeling only free-standing graphene whereas, experimentally, the graphene is on a SiC substrate and so electron absorption can occur within the substrate itself. The inelastic effects are even larger for the oscillations in the 14-22 eV range; these just barely visible in Fig. 1(a), in agreement with experimental results of Hibino et al.[9] Overall, the inclusion of inelastic effects greatly improves the agreement between theory and experiment for the LEER spectrum.

To further illustrate the inelastic effects, we display in Fig. 1(c) the computed dwell time for the electron wavepackets in the slab. The dwell times show marked increases at energies corresponding to the minima in the reflectivity. These energies correspond to the interlayer states; as discussed both by Merzbacher and in the previous Section, at these energies the electrons spend more time "bouncing to and fro" in the slab, producing attenuation in the reflectivity.[21] This attenuation extends out to neighboring energies through the mixing of the states in the slab, as expressed by Eq. (2). We note that the peak values of dwell time in Fig. 1(c), about 3 fs, are moderate in size compared to other systems discussed below. For graphene on Cu(111) we find much smaller dwell times, whereas for h-BN on various metals we find much larger ones. In all cases, the magnitudes of the dwell times are correlated with the widths of the relevant resonant states.

Figure 3 shows the relative insensitivity of our model to variations in the parameters. First in Fig. 3(a) we vary the inner potential, using values of $V_0 = 1$ eV or 50 eV rather than the nominal 18.1 eV. We find that the results are extremely insensitive to the $V_0$ value; even using $V_0 = 0$ eV produces nearly the same result except at energies < 1 eV where a noticeable drop in reflectivity occurs. In practice, an inner potential value of ≈10 eV can be used in our model, for all materials, with negligible error in the results. In Fig. 3(b) we illustrate the effect of varying the position of the edge of the slab, using values of 0.35 and 1.4 Å to the left of the graphene plane rather than the nominal 0.7 Å. A fairly small influence on the final result is found (use of the 0.35 Å values causes the dwell times to shift ≈0.02 fs downwards, whereas the 1.4 Å value shifts them ≈0.05 fs upwards). Finally in Fig. 3(c) we illustrate the influence of changing the magnitude of the imaginary part of the potential. Significant changes do now occur in the spectrum, as expected. All LEED theories to date use the imaginary part of the potential as a fitting parameter, chosen to match experiment. The value employed here of $V_i = 0.4 \, \text{eV} + 0.06 E$, based in part on the work of Kravoskii et al.,[18,19,20] is found to provide at least a semi-quantitative description of spectra for many different materials. Fine-tuning of these values can be done in specific cases to produce improved agreement if necessary.

**B. Graphene on Copper**

Figures 4(a) – 4(d) display computational results for a single layer of graphene on Cu(111), with separation of 3.58 Å between the graphene and the Cu. The red dashed line of Fig. 4(a) shows the computed reflectivity in the absence of inelastic effects, as in Ref. [17]. For reference, the band structures in the (0001) direction of graphite and in the (111) direction for fcc Cu are shown above that. Just as for the reflectivity of free-standing graphene (Fig. 1), interlayer states can form in accordance with the dispersive graphite band between about 0 and 6 eV in Fig. 4(b). The state in the present case will exist between the graphene and the Cu substrate, with energy that varies inversely with the separation between the graphene and the Cu.[17] For separation of 3.58 Å this interlayer state produces the minimum near 4.5 eV in Fig. 4(a). The computation of reflectivity is similar to that for a free-standing slab, except that the reflectivity in this case is obtained by matching wavefunctions in the center of the slab with those obtained from a bulk computation for Cu rather than by matching to vacuum states on either side of the slab.[17] There is a nearly free-electron (NFE) band of Cu located in the relevant energy range, as displayed in Fig. 4(c), with the bottom of this band being *below* the vacuum level for the (111) direction in Cu. A phase angle $\phi$ is obtained from the reflectivity, leading to the dwell time for a wavepacket in the slab and concomitant inelastic effects, using the method described in Section II(B).

The dwell time for graphene on the Cu(111) substrate is shown in Fig. 4(d). In this case there is very little effect of the interlayer state, since this state is actually quite broad and therefore does not induce any significant increase in dwell time. Values in Fig. 4(d) are all close to zero, with some being slightly below zero. In the latter case, *negative* dwell times would in principle lead to an *increase* in reflectivity according to $\exp(-\Delta t/\tau)$ from Section II(B). However, such an increase does not make physical sense; rather, a negative dwell time simply indicates that the wavepacket is reflected from the slab at some location that is farther out from the surface then where we placed our $-z_E$ location. We therefore truncate all of the dwell times at zero before applying the attenuation formula $\exp(-\Delta t/\tau)$. With the dwell times in Fig. 4(d) being all near zero, the inelastic effects on the reflectivity spectrum are small, as seen by the resultant solid blue line in Fig. 4(a).

Figures 4(e) – 4(h) show results for single layer graphene on Cu(001). The situation is different than for Cu(111), since in the (001) direction the onset of the NFE band occurs *above* the vacuum level, at 2.5 eV as seen in Fig. 4(g). Below this value there is an *energy gap* within which there are no propagating states in the Cu substrate and hence the reflectivity is simply $R=1$. However, to incorporate inelastic effects we still require values for the phase angle $\phi$ at these energies, and these must be obtained by some new method. Our technique for handling this situation is described in Appendix II. Essentially, we perform a computation for the graphene-Cu-graphene slab system, treating it as a free-standing slab although with the phase angle obtained in a slightly modified manner. With those values of phase in the band gap region, we can then construct a complete $\phi(E)$ curve and obtain dwell times as described in Section II(B). The results for graphene on Cu(001) are shown in Fig. 4(h). As for the case of graphene on Cu(111) the dwell times are again quite small, < 1 fs, with positive times obtained for energies just below on the onset of the NFE band and negative times just above the onset of the NFE band. The resulting reflectivity including inelastic effects forms a smooth curve through this energy region, as seen by the solid blue line in Fig. 4(e).

Experimental results are displayed in Ref. [17] for graphene on Cu(111) and (001). For the former, the observed reflectivity minimum is quite broad, in agreement with our theoretical result including inelastic effects. For the latter case, the smooth variation in reflectivity through the onset of the NFE band is also in general agreement with experiment, although the absolute magnitude of the reflectivity in the experiment is somewhat smaller than obtained in the theory. However, those experiments were obtained from samples that had been exposed to air for times ranging from several days to several weeks, and the reflectivity in the band gap region showed a significant decrease with the exposure time.[17] Oxidation of the Cu surface is thus seen to modify the spectrum. Future acquisition of a spectrum from a non-air-exposed surface is needed in order to achieve a more detailed comparison between experiment and theory.

### C. Hexagonal boron nitride on Cobalt

For the situations discussed above, the inelastic effects have been moderate or small, with dwell times of a few fs or less. We now turn to the case of h-BN on metal substrates such as Co or Ni, for which the inelastic effects are found to be quite large, with dwell times of 10's of fs or more. Reflectivities that are near unity in the absence of inelastic effects can thus be attenuated to < 0.1 when nearby bands with long dwell times are present. Thus, inclusion of the inelastic effects becomes quite important in the interpretation of the spectra. Experimental LEER spectra for h-BN on Co have been previously presented by Orofeo et al.,[27] and these were qualitatively compared with a computed h-BN band structure. We present here theoretically obtained LEER spectra, from which a more detailed interpretation of all the various spectral features can be made.

Separately, we have performed detailed experimental and theoretical LEER studies for h-BN on Ni.[28] One important result of those studies is that oxidation of the Ni surface (due to air exposure for at least several days between sample growth and introduction into the LEEM chamber) plays an important role in the experimental LEER spectra. LEED measurements have directly revealed the presence of this oxide.[28] The oxide produces a dipole at the interface, thus increasing the work function of the surface and shifting the onset of the Ni NFE bands from a location about 3 eV above the vacuum level to a location slightly *below* the vacuum level; this shift, in turn, significantly impacts the predicted LEER spectra. For our theoretical results of h-BN on Co, we similarly employ an oxide at the interface.

Figure 5 displays computed LEER spectra for h-BN on the oxidized Co(0001) surface. The computations employ a 5-layer hcp Co slab, together with oxygen and h-BN layers on either side. The O-Co separation is chosen to be 1.1 Å, obtained from a first-principles relaxation of a bare O layer on the Co surface. The BN-O separation is taken to be 1.68 Å, chosen in order to approximately match the theory with the experimental results of Orofeo et al.[27] For this value of BN-O separation the reflectivity minimum for a single h-BN layer occurs at about 7.5 eV, compared to 6.5 eV in the experiment. With a subsequent h-BN layer, a minimum appears near 2 eV, comparable to the experiment. In the theoretical spectra of Fig. 5 we show results for both the majority and minority spin, with onsets of the Co NFE bands being at 1.0 and 2.0 eV above the vacuum level, respectively. The average of the spin-resolved reflectivity curves, for each thickness of h-BN, can be compared to the experimental spectra (which are not spin resolved).[27]

Good agreement is obtained between the theoretical spectra and the experimental results of Orofeo et al.[27] Just as for the case of graphene, interlayer states form between the h-BN layers, with $n-1$ of these states forming for $n$ layers of h-BN.[12,27] There are three relevant bands of the h-BN that occur over the energy range shown in Fig. 5, and for each of these bands there will, in principle, be $n-1$ reflectivity minima. These $n-1$ bands are seen for the elastic results for the highest h-BN band in Fig. 5, although for the middle band they are not well resolved due to the energy-resolution of the computations. For the lowest band some of the minima are cut off in the elastic results due to the fact that the onsets of the NFE band lie above the vacuum level, but they are nonetheless clearly seen in the inelastic results. For the case of a single layer of h-BN, the relatively small BN-O separation precludes the presence of a well-defined interlayer state between those layers, with the resulting elastic spectrum displaying a broad minimum near 9 eV and a maximum near 7 eV.

Importantly, inclusion of inelastic effects has a profound influence on the spectra for all the h-BN thicknesses. In panel (e) of Fig. 6 we display dwell times for the case of 4 h-BN monolayers. Relatively large dwell times are found, particularly for the band centered at 6.8 eV. These dwell times produce low reflectivity for that band, and importantly, due to the mixing between nearby eigenstates that we employ in our model, this attenuation is then spread out to neighboring eigenstates that have high reflectivity in the absence of inelastic effects. Whereas a maximum occur in the elastic spectra at about 7 eV (for 1 ML h-BN) or 8 eV (for 2 – 4 ML h-BN), the inelastic effects produce sufficiently strong attenuation such that a *minimum* (for 1 ML) or very weak local maxima (for 2 – 4 ML) are produced in the spectra at these energies. Our interpretation of the spectra is in agreement with that presented previously by Orofeo et al.,[27] with our predicted reflectivity curves enabling a much more detailed understanding of the various spectral features. Again, proper treatment of inelastic effects is essential in producing spectra that can be compared to experiment.

**Summary**

In summary, we have presented a model for including inelastic effects into our computational methodology for low-energy electron reflectivity from surfaces. The model contains two components, one of which is rigorous (the dwell time for an electron in a slab) and the other approximate (the mixing between single-particle states in the slab). Our model is therefore not rigorously defined, both because of its approximate component and due to the way in which we have split the problem into two separable parts. However, we find from applying the model to a range of situations that we obtain results which are in reasonably good agreement with experiment. Additionally, the model can be easily incorporated into our reflectivity analysis method (i.e. using many of the same numerical quantities that are available in that procedure), so in this way it represents a useful advancement in the methodology. As illustrated in this work, inclusion of the inelastic effects permits detailed comparison between experimental and theoretical LEER spectra, from which structural parameters for the surface under study can be deduced.

**Acknowledgements**

This work was supported by the National Science Foundation under grant DMR-1205275, and the Office of Naval Research under MURI Grant No. N00014-11-1-0678.

**Appendix I**

In this Appendix we provide details of an improved method for dealing with the "mixed" states discussed in Ref. [12] that act to couple slabs in adjoining simulation cells. To introduce this topic, we first review the basic properties of reflected or diffracted waves as described in Ref. [12]. We initially consider a slab of material with a semi-infinite expanse of vacuum on either side. In that case, denoting the direction perpendicular to the slab surface by $z$, then the wavefunction in the vacuum for a propagating state with energy $E$ will consist of a travelling wave $\exp(i\kappa_\mathbf{g} z)$ multiplied by a sum of lateral waves of the form $A_\mathbf{g} \exp[i(g_x x + g_y y)]$. Here, the lateral wavevector is $\mathbf{g} = (g_x, g_y)$, $A_\mathbf{g}$ is an amplitude, $\kappa_\mathbf{g} = \sqrt{2m(E-E_V)/\hbar^2 - g_x^2 - g_y^2}$ is the $z$-component of the wavevector in the vacuum, $E_V$ is the vacuum energy, and $m$ is the free-electron mass. Diffracted beams have $(g_x, g_y) \neq (0,0)$; these exist only for $E - E_V \geq \hbar^2(g_x^2 + g_y^2)/2m$. Of course, at lower energies, evanescent states with substantial $(g_x, g_y) \neq (0,0)$ character exist; they extend out from the slab surface. These states decay to zero as the distance into the vacuum approaches infinity, so the amplitude of their $(g_x, g_y) = (0,0)$ component far into the vacuum is zero.

Now let us turn to the periodically repeating vacuum-slab-vacuum geometry of our VASP computation. With the finite vacuum width, the evanescent states cannot possibly decay exactly to zero, and as a consequence, all such states acquire a small, nonzero $(g_x, g_y) = (0,0)$ component in the vacuum. In practice, the magnitude of this nonzero $(g_x, g_y) = (0,0)$ component can be much greater than what would arise e.g. simply from extrapolating the exponentially decaying part of the wavefunction out to the edge of the vacuum region. Such states are said to have "mixed" character; they exist at energies less than $\hbar^2(g_x^2 + g_y^2)/2m$ and have primarily $(g_x, g_y) \neq (0,0)$ character, but they do not decay to zero in the vacuum. Examples of such states are shown in Figs. 3(a) and 3(b) of Ref. [16]. For these *mixed* states, we find that the magnitude of their $(g_x, g_y) = (0,0)$ component in the vacuum varies with the vacuum width used in the simulation. In this sense the states are spurious, their $(g_x, g_y) = (0,0)$ component is an artifact of the finite cell size used in the simulation. Importantly, the nonzero $(g_x, g_y) = (0,0)$ character of these states will, if used in our reflectivity analysis, lead to some reflectivity associated with these states. Those reflectivity values are themselves artifacts, and hence these mixed evanescent states must somehow be rejected from the analysis.

In order to identify the mixed states, we have in our prior work employed the quantity $\sigma$ defined in the Supplemental Material of Ref. [12], which is the overlap in the vacuum between the wavefunction of a state and a simple oscillatory wavefunction expected for a free electron states. Values of $\sigma$ are generally small ($\lesssim 10^{-2}$) for spurious mixed states and large ($\approx 1$) for *bona fide* propagating states. Importantly, since our definition of $\sigma$ includes a scale factor relating to the width of the vacuum region, then for the *bona fide* propagating states, their $\sigma$ values necessarily approach unity for sufficiently large vacuum width (this point is further discussed at the end of this Appendix). However, for certain mixed states we occasionally find $\sigma$ values of 0.1 or greater, so separating them from the *bona fide* states can become problematic. In our previous work we employed a discriminator value of $\sigma = 0.8$, rejecting all states with smaller $\sigma$ values.[12] Although that value worked reasonably well for the multi-layer graphene case, we find in other cases that this discriminator value can sometimes lead to rejection of *bona fide* propagating states, something that is important to avoid especially when inelastic effects are included. We have therefore employed in the present analysis a smaller discriminator value, $\sigma = 0.1$, but we supplement that by detailed inspection of the results, including their dependence on the vacuum width of the simulation. In this way, we can further reject the occasional state that is identified as having mixed character but which nevertheless has a $\sigma$ value of 0.1 or greater.

To illustrate this procedure, we display in Fig. 6 a small energy window for a simulation involving a free-standing slab of 4 graphene layers, with three different vacuum widths corresponding to total simulation cell widths of 4.0266, 5.3688, and 6.711 nm (the vacuum width is given by these values minus the width of the 4-layer graphene slab, 1.0 nm). We choose the energy window to correspond to the location of the narrow bulk graphite band centered at about 7.25 eV as seen in the band structures of Figs. 1 and 4. We plot the energy of eigenstates in the respective slab computations in Figs. 6(a) – (c), with the narrow bulk band shown for reference in Fig. 6(d). The eigenstates in Figs. 6(a) – (c) are separated into bands, labeled 1 – 5 in each panel. It is clear that there is one dispersive band, e.g. labelled 1 in Fig. 6(a), and four nearly flat bands. The dispersive band is associated with a propagating states in the vacuum; it moves up in energy as the simulation width increases, simply reflecting the location of the allowed energy window for propagating states of our periodic vacuum-slab-vacuum system. In each of Figs. 6(b) and 6(c), the dispersive band crosses a flat band, resulting in band anti-crossing behavior.

The states associated with the flat bands (or flat portions of bands) in Fig. 6 are all of the "mixed" type defined above, that is, they have large, in-plane oscillatory nature in the graphene planes, with very small (but constant, as a function of *z*) $(g_x, g_y) = (0,0)$ component far out in the vacuum. In particular, for Fig. 6(a), the wavefunctions of bands 2 and 3 at the midpoint of the wavevector range are identical to those shown in Figs. 3(a) and 3(b) of Ref. [16] (the energies in the present computation have been updated slightly compared to those of Ref. [16], but nevertheless the wavefunctions shown in Ref. [16] are identical to those of the present computation).[29] In contrast, the wavefunction associated with the dispersive band in Fig. 6(a) has character more like Fig. 3(c) of Ref. [16], i.e. with substantial $(g_x, g_y) = (0,0)$ component both in the vacuum and in the slab.

In Figure 7 we plot the $\sigma$ values for all the slab states associated with the computations of Fig. 6, using a wide energy range of 0 – 20 eV but with expanded view of the 6.9 – 7.4 eV range of Fig. 6. Clearly, two ranges of $\sigma$ values are dominant: near unity (which are the *bona fide* propagating states), and $\approx 10^{-2}$ (which are all spurious mixed states). However, a few intermediate values also occur, and furthermore, and these intermediate $\sigma$ values are not associated with a state at a fixed energy as we vary the simulation cell width. By comparison of Figs. 6 and 7, the intermediate $\sigma$ values lying between 0.1 and about 0.8 can be seen to occur when the energy of the dispersive band is near, or crossing, the energy of a flat band. At such band crossings (anti-crossings), interaction between the states in the bands leads to larger $\sigma$ values for the mixed states. Similar band crossings account for the intermediate $\sigma$ values at all other energies in Fig. 7.

We therefore reject states that have intermediate $\sigma$ values resulting from this interaction between the mixed and propagating states. For example, band 2 in Figs. 6(a) and 7(a) acquires some increased $\sigma$ value due to its proximity to band 1. Hence, all states in band 2 are rejected (even though one state in that band has a $\sigma$ value of $> 0.1$). The situation for the band anti-crossings in Figs. 6(b) and 7(b) and in Figs. 6(c) and 7(c) is slightly more complicated. We could simply reject all of the states in bands 1 and 2 for the former case and bands 2 and 3 for the latter, which would certainly eliminate all spurious states, but we would then end up eliminating a few *bona fide* states as well (i.e. on the dispersing ends of the respective bands). Thus, we choose to reject only those portions of the bands with states having $\sigma$ values of $< 0.1$, along with a few nearby states in each band that have $\sigma$ values between 0.1 and about 0.8.

The situation just described for multi-layer graphene slabs is actually a relatively easy one, since the $\sigma$-separation between propagating and evanescent states is straightforward to discern and the mixed states can be readily dealt with. In general, the ease with which this analysis can be conducted depends on both the nature of the states in the slab and the width of the vacuum region in the simulation. If we were to display in Fig. 7 results for the multi-layer graphene for smaller simulation cell width (e.g. $\approx$2.5 nm), then the separation of the *bona fide* and mixed states would be much less clear; in particular, many of the *bona fide* states would have $\sigma$ values significantly less than unity. Again, since our definition of $\sigma$ includes a scale factor relating to the width of the vacuum region, then for *bona fide* states, their $\sigma$ values always approach unity for sufficiently large vacuum width. For example, consider a state that has much of its wavefunction concentrated in the slab, but that nevertheless has a small (non-spurious) tail with $(g_x, g_y) = (0,0)$ character extending out into the vacuum. Such states do not occur for multi-layer graphene because of the nature of its states (i.e. the narrow bands of graphite shown in Figs. 1 and 4 all have in-plane oscillatory nature with essentially zero $(g_x, g_y) = (0,0)$ character), but for h-BN, with its inequivalent B and N atoms, such states *do* occur (and they *do* contribute to the reflectivity) since some of its narrow bands acquire some nonzero $(g_x, g_y) = (0,0)$ character. How then do we distinguish between spurious and *bona fide* states in this case? The answer, as just stated, is that even these *bona fide* states with relatively small $(g_x, g_y) = (0,0)$ character in the vacuum have $\sigma$ values that approach unity for sufficiently large vacuum width. A unity value of $\sigma$, as a function of increasing vacuum width, implies that the ratio of the wavefunction magnitude in the vacuum compared to that within the slab is constant, i.e. as expected when the propagating part of the wavefunction in the vacuum is a true, invariant feature. Conversely, if a state maintains a low $\sigma$ value, e.g. $< 0.1$, as a function of increasing vacuum width, then that implies that its wavefunction magnitude in the vacuum is actually *decreasing* relative to that within the slab. This varying ratio of wavefunction magnitude in the vacuum compared to within the slab (i.e. an overall decrease as a function of increasing vacuum width, together with fluctuations due to the presence of nearby propagating states) is the hallmark of a spurious, mixed state.

**Appendix II**

In this Appendix we provide details of the method used to deduce scattering phase shifts for an overlayer on a substrate for the situation when the energy of the incident electron falls within a band gap for the electronic states of the substrate. An important distinction should be made between the approximate method that we employ for our reflectivity analysis compared to the more rigorous computation of Krasovskii et al.[18,19] Those authors use a *semi-infinite* model for the substrate, computing its states for both real and imaginary values of the wavevector $k_z$. Within a band gap region, no eigenstates exist for purely real values of $k_z$, but *evanescent states* corresponding to imaginary $k_z$ values do naturally occur. It is these evanescent states that will couple to an incident electron wave that has propagated through the overlayer, and the phase shift in the reflected wave relative to the incident one is determined by the evanescent states. In our approximate model, we do *not* compute these evanescent states for the bulk substrate. Rather, we first perform a supercell computation for a vacuum-overlayer-substrate-overlayer-vacuum system with a substrate of limited thickness (typically only a few atomic layers), and then we match states of that system to states of a periodic (infinite) bulk substrate with only real $k_z$ values. Thus, we do not obtain evanescent states of our bulk substrate and hence these are not available for the purpose of determining the phase of the reflected wave.

However, since we compute the spectrum of states for the vacuum-overlayer-substrate-overlayer-vacuum supercell, we *do* obtain some information on evanescent states of the bulk substrate. That is, even with purely real $k_z$ values, that slab computation will include states that exist for energies within the band gap of the bulk substrate. These states will vary exponentially with distance (decaying or growing) within the substrate portion of the slab, for energies in the band gap. From these states, we wish to construct a linear combination that has purely decaying exponential behavior in the substrate portion of the slab.

In our general methodology for reflectivity analysis of a free-standing slab,[12] we construct linear combinations of states that are even or odd relative to the center of the slab (assuming, for ease of discussion, a potential that is symmetric about this central point). In the vacuum region on the far left-hand side of the simulation cell these functions vary like $A_e \cos(kz + \delta_e)$ and $A_o \sin(kz + \delta_o)$, respectively, thus defining $\delta_e$ and $\delta_o$. On the far right-hand side of the simulation cell they vary like $A_e \cos(kz - \delta_e')$ and $A_o \sin(kz - \delta_o')$, thus defining $\delta_e'$ and $\delta_o'$, and with $\delta_e = \delta'_e$ and $\delta_o = \delta'_o$ for a symmetric potential. In our standard method, we form further linear combinations such that on the right-hand side of the slab there is only an outgoing wave, whereas on the left-hand side there are both incoming and outgoing waves. In that way, the transmission coefficient is found to be (see Supplemental Information of Ref. [12])

$$T = \left| \frac{2\cos(\delta'_e - \delta'_o)}{e^{i(\delta'_e + \delta_o)} + e^{i(\delta_e + \delta'_o)}} \right|^2. \quad (A1)$$

Examining this result, we see that in order to achieve $T = 0$ (i.e. reflectivity of $R = 1$), then we must have $\delta'_e - \delta'_o = \pi/2$ or $3\pi/2$, modulo $2\pi$. If this situation occurs for an energy lying within a band gap in the bulk electronic spectrum, then we would have achieved our goal of constructing a purely exponentially decaying state within the substrate portion of the slab (since its amplitude is clearly going to zero on the right-hand side of the slab).

However, for energies within a band gap we will not in general find $R = 1$ in our analysis of the free-standing slab. In these cases we still desire to minimize the amplitude of the wavefunction on the right-hand side of the slab. For this purpose we form a slightly different linear combination than that which was employed for obtaining Eq. (A1). Rather than demanding that the incoming wave on the right-hand side of the slab be zero, we instead desire to minimize the amplitude of both the incoming and outgoing waves on the right-hand side of the slab. Consider forming in the vacuum region on the right-hand side of the slab the combination

$$A_o A_e \cos(kz - \delta'_e) \pm A_e A_o \sin(kz - \delta'_o). \quad (A2)$$

When $\delta'_e - \delta'_o = \pi/2$ or $3\pi/2$ then this combination is zero when using the lower or upper sign, respectively. However, even when $\delta'_e - \delta'_o$ deviates slightly away from $\pi/2$ or $3\pi/2$ then the combination is still small. The reason that it is small is that, within the substrate portion of the slab, the linear combination (A2) approximately takes a form proportional to $A_o A_e \cosh(kz) - A_e A_o \sinh(kz)$,[30] leading to a purely exponentially decaying dependence for $z > 0$. Given this linear combination (A2), then the ratio of reflected to incident waves on the left-hand side of the slab is easily found to be $(e^{-i\delta_e} \pm e^{-i\delta_o})/(e^{i\delta_e} \mp e^{i\delta_o})$. The phase $\phi$ of this complex quantity then gives us the phase angle to use in our dwell time analysis, for energies within a bulk band gap.

As a test of the applicability of the linear combination (A2) for forming states on the right-hand side of the slab that have small amplitude, we consider the amount by which $\delta'_e - \delta'_o$ deviates away from $\pi/2$ or $3\pi/2$. For the cases considered in the body of this work with substrate thicknesses of 3 or 5 atomic layers we generally obtained deviations of less than 0.1, for which we consider the resulting $\phi$ values to be sufficiently accurate since they do not significantly change even if a thicker substrate portion of the slab is employed. In situations where the deviation is larger, then we redo the analysis using a thicker substrate portion of the slab, in order to decrease the deviation in $\delta'_e - \delta'_o$ away from $\pi/2$ or $3\pi/2$.

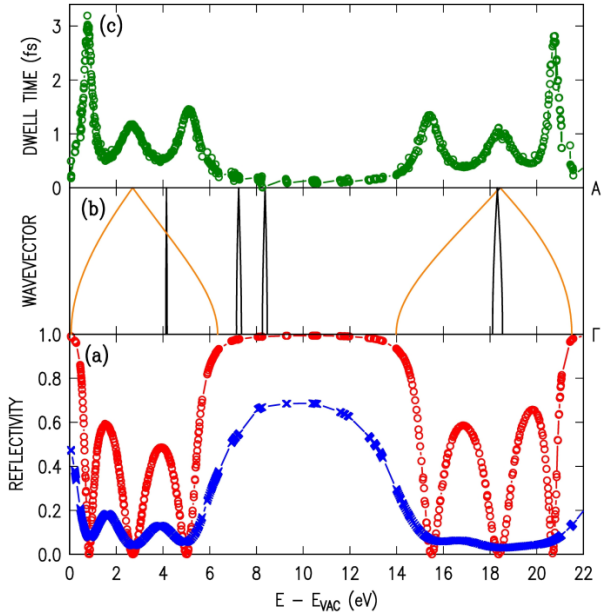

FIG 1. (Color on-line) Theoretical results for reflectivity of a free-standing slab of 4-layer graphene. (a) Reflectivity without (red circles) and with (blue x-marks) inelastic effects. (b) Band structure of graphite, in (0001) direction. (c) Dwell time for a reflected wavepacket in the slab, corresponding to the time difference between a quantum-mechanical particle compared to a classical particle reflecting off of the slab. Maxima in the dwell time correspond to resonant interlayer states of the slab, which in turn give rise to minima in the reflectivity.

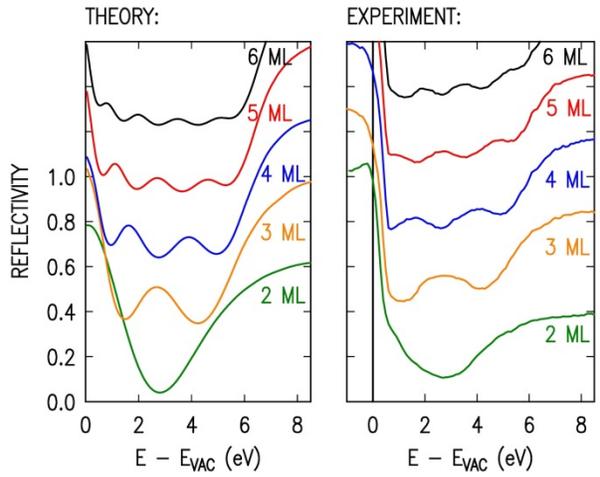

FIG 2. (Color on-line) Theoretical (left) and experimental (right) results for reflectivity of free-standing slab of multilayer graphene of various thickness in monolayers (ML) as indicated. The absolute reflectivity scale for the 2 ML case is shown on the left, with subsequent curves shifted upwards by 0.3 units each.

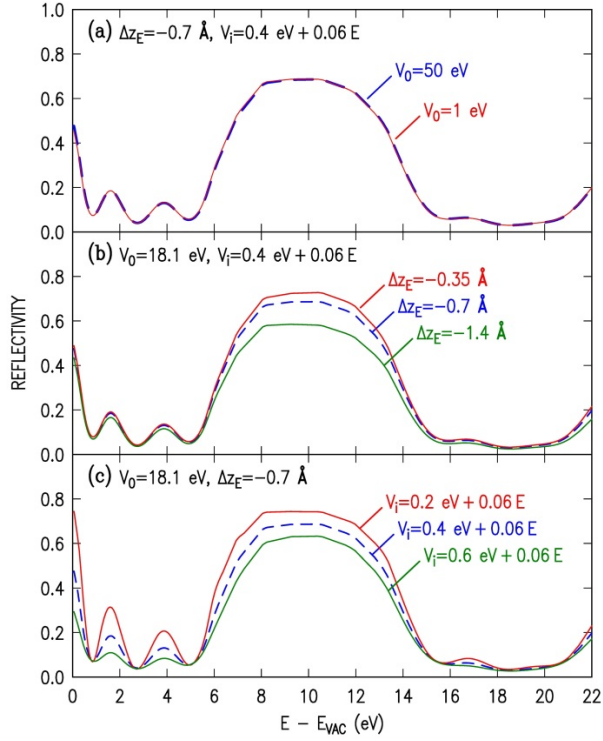

FIG 3. (Color on-line) Theoretical results for reflectivity of a free-standing slab of 4-layer graphene, illustrating the dependence on parameters in the computation: (a) variation of inner potential $V_0$, (b) variation in the location of the edge of the slab relative to the left-most plane of carbon atoms, $\Delta z_E$, and (c) variation in the imaginary portion of the potential.

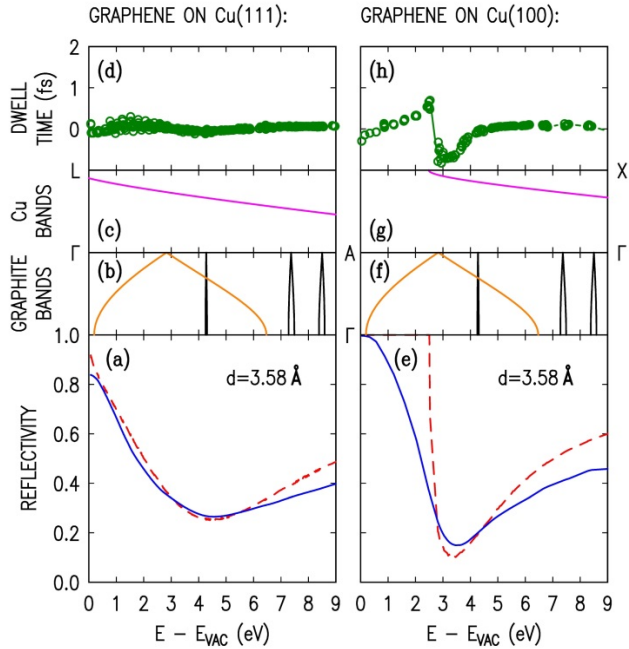

FIG 4. (Color on-line) Theoretical results for reflectivity of monolayer graphene on Cu(111) (left) and Cu(100) (right), with separation of 3.58 Å between the graphene and the Cu. (a) and (e) Reflectivity without (red dashed lines) and with (blue solid lines) inelastic effects. (b) and (f) Band structure of graphite, in (0001) direction. (c) and (g) Band structure of Cu, in (111) (left) or (100) (right) directions. (d) and (h) Dwell time for a reflected wavepacket in the graphene-substrate slab.

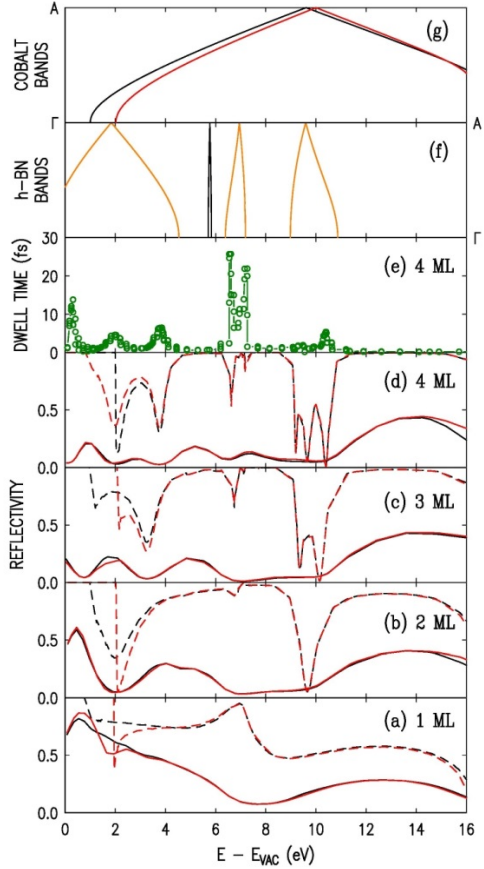

FIG 5. (Color on-line) (a)-(d) Theoretical results for reflectivity of 1-4 layer h-BN on oxidized Co (0001) surface, with separation of 1.1 Å between the O and Co, 1.68 Å between BN and O, and 3.3 Å between BN layers. Reflectivity without (dashed lines) and with (solid lines) inelastic effects. Black and red are for majority and minority spins respectively. (e) Dwell time for 4-layer h-BN on oxidized Co as in (d), majority spin. (f) h-BN band structure. (g) Co band structure. Black is majority spin, red is minority spin.

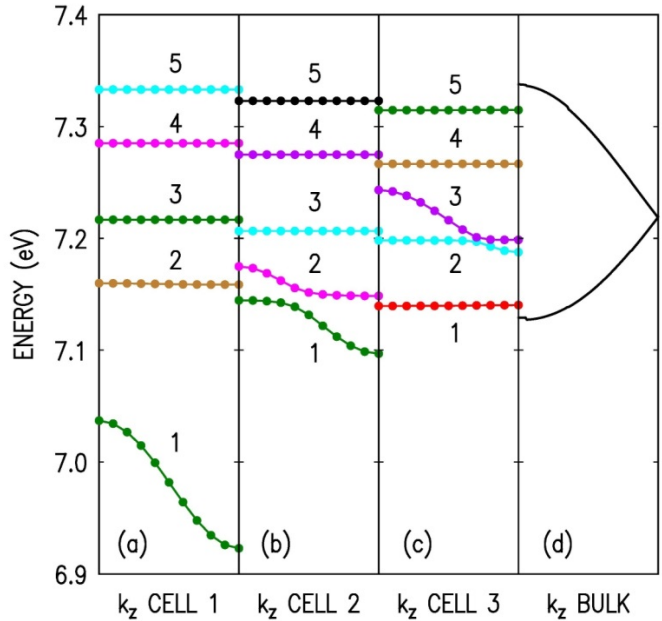

FIG 6. (Color on-line) Computed energy bands for 4 layers of graphene, with simulation cell widths of (a) 4.0266 nm, (b) 5.3688 nm, and (c) 6.711 nm. Energy bands of bulk graphite are shown in (d). The wavevector ($k_z$) ranges in (a) – (c) are given by $\pi$ divided by the cell width, whereas in (d) the wavevector varies from the Γ-point to the A-point. Energies in (a) – (c) are plotted relative to the vacuum level of each slab, whereas in (d) the bulk band is positioned by aligning the potential of the bulk with the potential of the 4-layer slab in (a). The colors of the bands in (a) – (c) is chosen to coincide with those of Fig. 7.

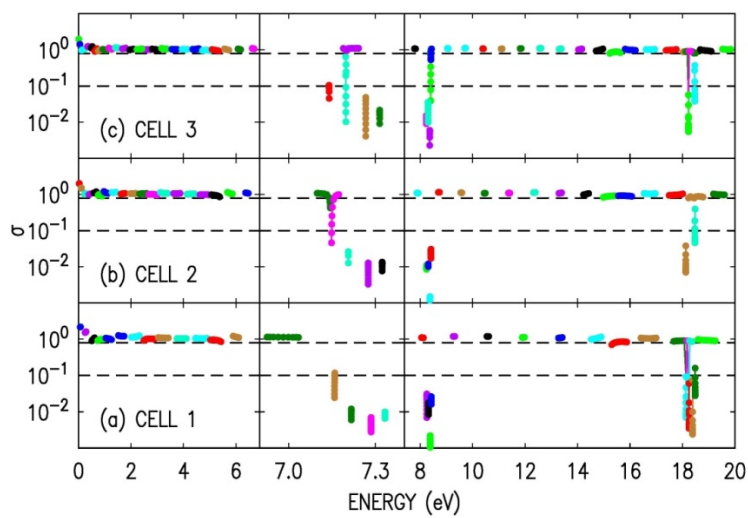

**FIG 7.** (Color on-line) Plot of the magnitude of the oscillatory component of the wavefunction in the vacuum, $\sigma$, as a function of the energy of the states, for the same simulation cells (a) – (c) used in Fig. 6. Different colors are used to represent different bands. Dashed lines are drawn in each panel at $\sigma$ values of 0.1 and 0.8. Note the expanded energy scale over 6.9 – 7.4 eV.

**References**

[1] F. Jona, J. Phys. C: Solid State Phys. **11**, 4271 (1978), and references therein.
[2] M. A. Van Hove, W. Moritz, H. Over, P. J. Rous, A. Wander, A. Barbieri, N. Materer, U. Starke, and G. A. Somorjai, Surf. Sci. Rep. **19**, 191 (1993).
[3] H. Lüth, *Surface and Interfaces of Solid Materials* (Springer; 3rd edition, 1995).
[4] R. Zdyb and E. Bauer, Phys. Rev. Lett. **88**, 166403 (2002).
[5] B. T. Jonker, N. C. Bartelt, and R. L. Park, Surf. Sci. **127**, 183 (1983).
[6] W. F. Chung, Y. J. Feng, H. C. Poon, C. T. Chan, S. Y. Tong, and M. S. Altman, Phys. Rev. Lett. **90**, 216105 (2003).
[7] M. Altman, J. Phys.: Condens. Matter **17**, S1305 (2005).
[8] H. Hibino, H. Kageshima, F. Maeda, M. Nagase, Y. Kobayashi, and H. Yamaguchi, Phys. Rev. B 77, 075413 (2008).
[9] H. Hibino, H. Kageshima, F. Maeda, M. Nagase, Y. Kobayashi, and H. Yamaguchi, e-J. Surf. Sci. Nanotech. **6**, 107 (2008).
[10] P. Sutter, J. T. Sadowski, and E. Sutter, Phys. Rev. B **80**, 245411 (2009).
[11] Luxmi, N. Srivastava, and R. M. Feenstra, J. Vac. Sci Technol. B **28**, C5C1 (2010).
[12] R. M. Feenstra, N. Srivastava, Q. Gao, M. Widom, B. Diaconescu, T. Ohta, G. L. Kellogg, J. T. Robinson, and I. V. Vlassiouk, Phys. Rev. B **87**, 041406(R) (2013), and references therein.
[13] G. Kresse and J. Hafner, Phys. Rev. B **47**, RC558 (1993).
[14] G. Kresse and J. Furthmuller, Phys. Rev. B **54**, 11169 (1996).
[15] G. Kresse, and D. Joubert, Phys. Rev. B **59**, 1758 (1999).

[16] R. M. Feenstra and M. Widom, Ultramicroscopy **130**, 101 (2013).

[17] N. Srivastava, Q. Gao, M. Widom, R. M. Feenstra, S. Nie, K. F. McCarty, I. V. Vlassiouk, Phys. Rev. B **87**, 245414 (2013).

[18] E. E. Krasovskii, W. Schattke, V. N. Strocov, and R. Claessen, Phys. Rev. B **66**, 235403 (2002).

[19] E. E. Krasovskii and V. N. Strocov, J. Phys.: Condens. Matter **21**, 314009 (2009).

[20] J. Ingo Flege, A. Meyer, J. Falta, and E. E. Krasovskii, Phys. Rev. B **84**, 115441 (2011).

[21] E. Merzbacher, *Quantum Mechanics* (Wiley; 2nd edition, 1970).

[22] D. R. Penn, J. Vac. Sci. Technol. **13**, 221 (1976), and references therein.

[23] A. Goswami and N. D. Lisgarten, J. Phys. C: Solid State Phys. **15**, 4217 (1982), and references therein.

[24] S. Y. Zhou, G.-H. Gweon, and A. Lanzara, Ann. Phys. **321**, 1730 (2006).

[25] C. Kittel, *Introduction to Solid State Physics* (Wiley; 8th edition, 2005).

[26] M. Posternak, A. Baldereschi, A. J. Freeman, E. Wimmer, and M. Weinert, Phys. Rev. Lett. **50**, 761 (1983).

[27] C. M. Orofeo, S. Suzuki, H. Kageshima, and H. Hibino, Nano Research **6**, 335 (2013).

[28] P. C. Mende, Q. Gao, M. Widom, R. M. Feenstra, to be published.

[29] The band structure displayed in Fig. 2 of Ref. [16] is one that was inadvertently computed with only 4-5 significant figures used to specify some of the atomic positions in the unit cell. Nevertheless, all other results in that work were computed with a full 6-7 significant figures used for all atomic positions. The energy eigenvalues of 3.24, 7.09, and 7.14 eV specified in Ref. [16] are shifted to 3.24, 7.16, and 7.22 eV for the higher precision eigenvalues of the present work.

[30] More generally, the linear combination (A2) in the substrate portion of the slab would contain $\cosh(z)$ and $\sinh(z)$ terms as *envelope functions* of the complete wavefunction, but this does not change our argument.